\documentclass[twocolumn,amsmath,amssymb,floatfix,superscriptaddress,prb]{revtex4-2}
\usepackage{url,epsf}
\usepackage{comment}
\usepackage{appendix}
\usepackage{physics}
\usepackage{multirow}
\usepackage[colorlinks=true,linkcolor=blue]{hyperref}
\usepackage{amsmath,bm,graphicx,epsfig}
\usepackage{xcolor}
\usepackage[english]{babel}
\usepackage{amsfonts,stmaryrd,amssymb}
\usepackage{enumerate} 

\newcommand{\Kelvin}{\text{K}}

\newcommand{\meV}{\text{meV}}
\newcommand{\eV}{\text{eV}}

\newcommand{\bQ}{\mathbf{Q}}
\newcommand{\bG}{\mathbf{G}}
\newcommand{\bK}{\mathbf{K}}
\newcommand{\bq}{\mathbf{q}}
\newcommand{\bk}{\mathbf{k}}
\newcommand{\bp}{\mathbf{p}}

\newcommand{\bz}{\mathbf{z}}
\newcommand{\br}{\mathbf{r}}
\newcommand{\eqn}[1]{(\ref{#1})}
\newcommand{\mr}{moir\'e~}

\begin{document}
\title{Designer Meron Lattice on the Surface of a Topological Insulator}

\author{Daniele Guerci}
\affiliation{Center for Computational Quantum Physics, Flatiron Institute, New York, New York 10010, USA}

\author{Jie Wang}
\affiliation{Center for Computational Quantum Physics, Flatiron Institute, New York, New York 10010, USA}

\author{J. H. Pixley}
\affiliation{Department of Physics and Astronomy, Center for Materials Theory, Rutgers University, Piscataway, New Jersey 08854, USA}
\affiliation{Center for Computational Quantum Physics, Flatiron Institute, New York, New York 10010, USA}

\author{Jennifer Cano}
\affiliation{Department of Physics and Astronomy, Stony Brook University, Stony Brook, New York 11974, USA}
\affiliation{Center for Computational Quantum Physics, Flatiron Institute, New York, New York 10010, USA}

\begin{abstract}
We present a promising route to realize spontaneous magnetic order on the surface of a 3D topological insulator by applying a superlattice potential.
The superlattice potential lowers the symmetry of the surface states and creates tunable van Hove singularities, which, when combined with strong spin-orbit coupling and Coulomb repulsion give rise to a topological meron lattice spin texture. The periodicity of this designer meron lattice can be tuned by varying the periodicity of the superlattice potential.
We employ Ginzburg-Landau theory to classify the different magnetic orders and show that the magnetic transition temperature reaches experimentally accessible values. 
Our work introduces a new direction to realize exotic quantum order by engineering interacting Dirac electrons in a superlattice potential, with promising applications to spintronics.
\end{abstract}

\maketitle

\section{Introduction}

Three-dimensional topological insulators (TIs) host gapless surface Dirac cones protected by time reversal symmetry~\cite{Fu2007,Moore2007,Roy2009,Konig_2007,LF_2009,hsieh2009topological,Hsieh2009ATT,science.1173034,Xia_2009,PhysRevLett.106.126803,PhysRevLett.106.216803}. 
In the presence of strong interactions, time-reversal symmetry can be spontaneously broken, gapping the Dirac cone.
The result is an exotic magnetically ordered surface exhibiting the quantized anomalous Hall effect~\cite{LF_2009,Wu_Si_2011,2012_1_AStern,2012_2_AStern,PhysRevB.86.155146,PhysRevB.86.161110,PhysRevB.88.205107,PhysRevB.91.155405}. However, 
for all measured TIs, the Coulomb interaction is too weak to induce the magnetically ordered phase \cite{LF_2009,Wu_Si_2011,2012_2_AStern,PhysRevB.91.155405}. 

\begin{figure}
    \centering
    \includegraphics[width=0.4\textwidth]{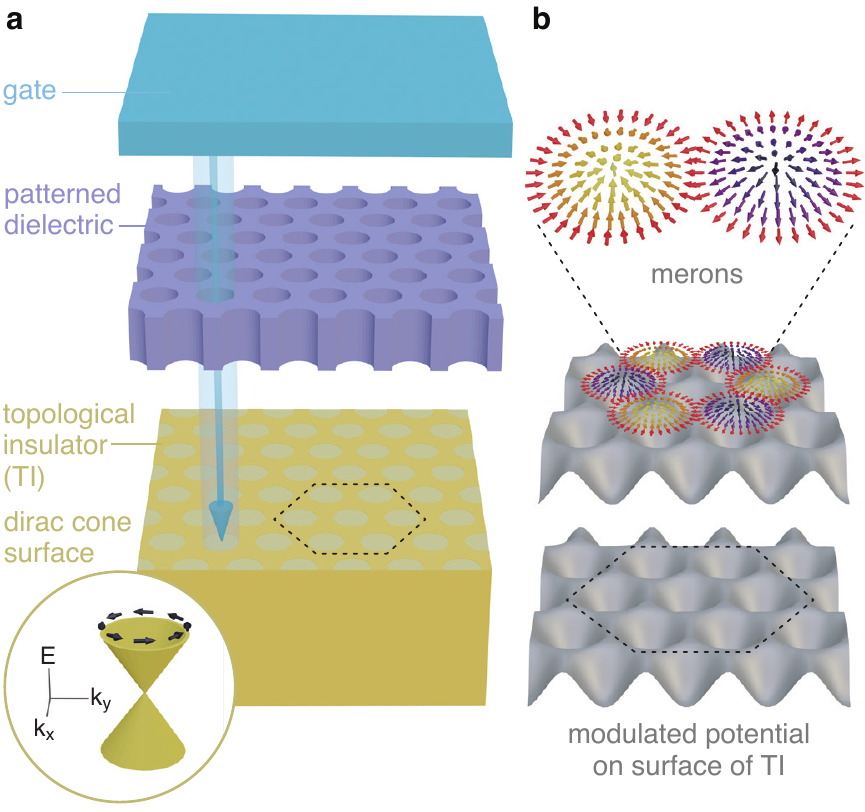}
    \vspace{-0.3cm}
    \caption{\textbf{Schematic of experiment and meron spin texture.} (a) The experiment consists, from top to bottom, of a metallic gate, a patterned dielectric and a topological insulator. Applying a bias between the metallic gate and the topological insulator imposes a modulated potential on the surface of the TI. (b) Magnetization exhibiting half-integer winding characteristic of a meron spin texture. Merons pinned by the superlattice potential form a meron lattice.}
    \label{Sketch_exp}
\end{figure}

In this manuscript, we show that a superlattice potential enhances correlation effects and provides a new and experimentally accessible route to realize spontaneous magnetic order on the surface of a TI as depicted in Fig.~\ref{Sketch_exp}(a). 
The superlattice potential downfolds and strongly renormalizes the low-energy band structure, creating satellite Dirac cones without opening a gap~\cite{PhysRevB.103.155157} and inducing strong van Hove singularities (VHSs)~\cite{PhysRevX.11.021024}. 
We show that the superlattice induced VHSs drive a spin density wave instability that results in an exotic meron lattice taking place at arbitrarily small values of the electron-electron interaction.
A meron, as shown in Fig.~\ref{Sketch_exp}(b), is topologically equivalent to half a skyrmion: magnetic moments in its core point up or down, but magnetic moments along its boundary are in-plane~\cite{Gao_2019}. 
While topological spin textures such as skyrmions have been observed in non-centrosymmetric magnets~\cite{Rosch2009,Pappas_2009,PhysRevLett.102.186602,Nagaosa_2013}, ultrathin magnetic films~\cite{Heinze_2011,Romming_2013,Gross_2018} and multiferroic insulators~\cite{Seki_2012,Yu_2012}, 
a meron lattice has only been observed recently~\cite{Yu_2018,Gao2020}.

The meron lattice we describe on the surface of the TI is stabilized by the interplay between the superlattice potential and the strong spin-orbit coupling (SOC) on the TI surface that ``locks'' the spin to the momentum, forcing the magnetic moments to rotate in space about an in-plane axis.
It features several novel aspects: (1) it leaves the surface Dirac cone gapless; (2) each unit cell exhibits two merons with opposite topological charge;
and (3) the meron lattice periodicity is determined by the applied potential, {\it i.e.}, it can be chosen by design. 
These features make it distinct from other magnetic textures induced on~\cite{LF_2009,Wu_Si_2011,2012_1_AStern,2012_2_AStern,PhysRevB.86.155146,PhysRevB.86.161110,PhysRevB.88.205107,PhysRevB.91.155405} or proximity-coupled~\cite{divic2021magnetic,Nomura2010,Hurst2015,Tiwari2019,Paul_2021} to the surface of a TI.
 
In this work we focus on the experimental setup shown in Fig.~\ref{Sketch_exp}(a), which depicts
a superlattice potential imposed on the surface of a TI by gating a patterned dielectric stacked above the TI surface. This approach was introduced to realize a superlattice potential on graphene with periodicity down to 35nm and strength $\sim 50$meV~\cite{Forsythe_2018,li2021anisotropic}. It offers great tunability: the periodicity, strength, and symmetry of the potential can be engineered. 
A superlattice potential has been studied theoretically to band engineer topological materials~\cite{Song_2010,PhysRevB.103.155157,PhysRevX.11.021024,Chou_2021,liu2021magnetic,shi2019gate,PhysRevB.103.155157,PhysRevX.11.021024,Rossi_2014,Schouteden_2016,Vargas_2017,Hennighausen_2019_1,Hennighausen_2019,dunbrack2021magic}. Strain has also been employed to modify the dispersion of surface states~\cite{Tang_2014,liu2014tuning}. 

\section{Model of an interacting TI surface}

\begin{figure}[]
\centering
\includegraphics[width=0.47\textwidth]{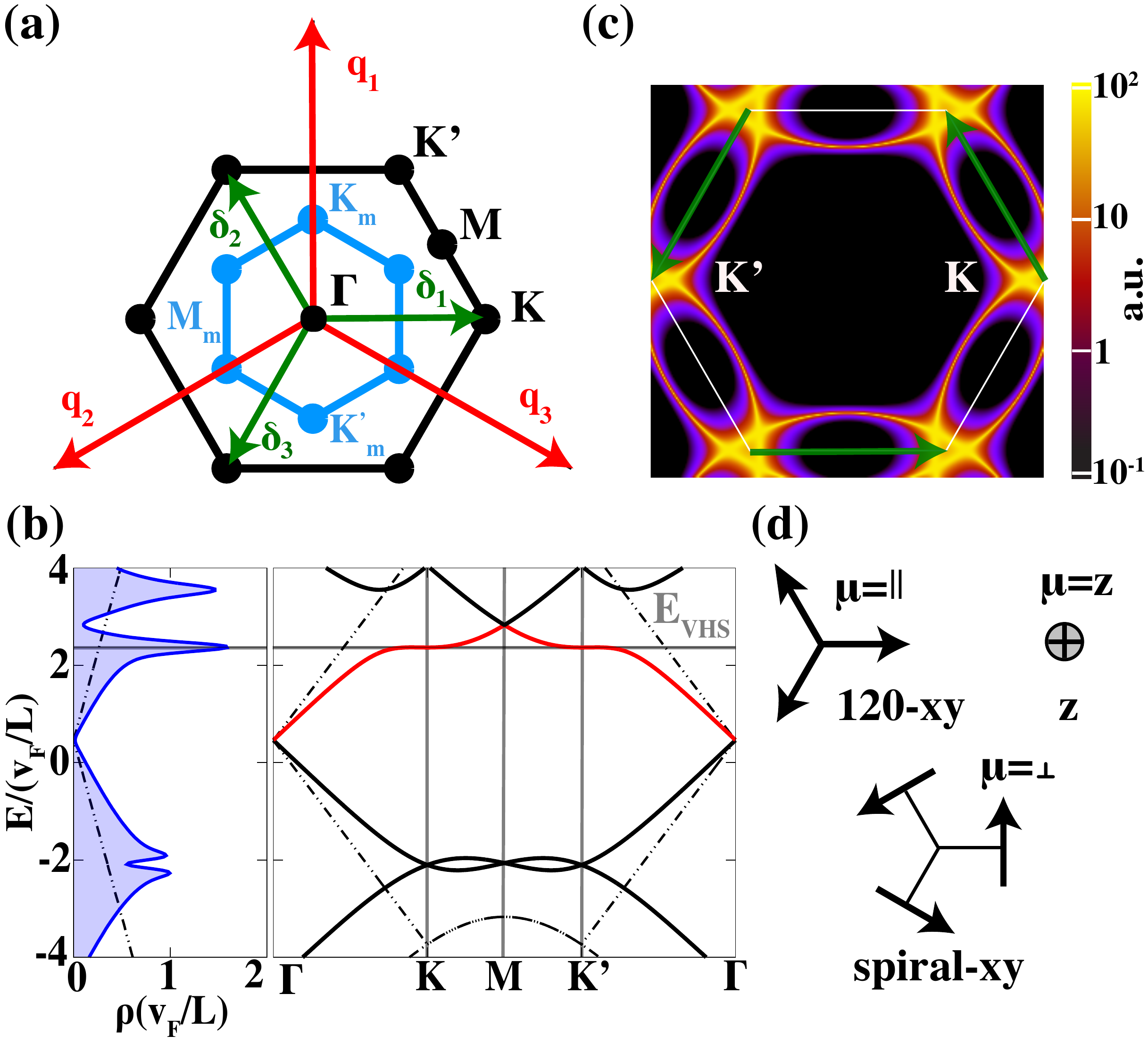}
\vspace{-0.3cm}
\caption{\textbf{Superlattice Brillouin zone, electronic properties of surface states and magnetic ordering basis}. (a) Black and cyan lines show the superlattice and magnetic Brillouin zones, respectively. The red and green arrows represent the wave vectors $\bq_j$ and the nesting vectors, $\bm\delta_j$, respectively. (b) Density of states and electronic dispersion for $w/(v_F/L)\simeq1.54$ ($L=20$nm, $w=20$meV). Dashed lines show the density of states and the band structure of the TI Dirac cone folded into the superlattice Brillouin zone. (c) Momentum resolved spectral function displaying the Fermi surface at the energy of the VHS with $w/(v_F/L)\simeq1.54$. (d) Decomposition of the order parameter into $\bm v_{\mu j}$.
}
\label{band_structure}
\end{figure}

We consider interacting electrons on the surface of a TI described by the Hamiltonian $\hat H=\hat H_0+\hat H_{\rm int}$. The non-interacting Hamiltonian $\hat H_0=\int d^2\br\, \hat\Psi^\dagger(\br)H_0(\br)\hat\Psi(\br)$ describes a spin-momentum locked Dirac cone subject to a superlattice potential, 
\begin{equation}
\label{model_Hamiltonian}
    H_0(\br)=v_F\left(-i\bm\nabla_\br\times\bm\sigma\right)_z+\sigma_0\,w(\br),
\end{equation}
where $\bm\nabla_\br=(\partial_x,\partial_y)$, $\bm\sigma=(\sigma_x,\sigma_y,\sigma_z)$ are the Pauli matrices, $\sigma_0$ the identity, and $w(\br)=2w\sum_{j=1}^3\cos\left(\bq_j\cdot\br\right)$ is the hexagonal superlattice potential, with amplitude $2w$ and wave vectors $\bq_{1,2,3}$ illustrated as red vectors in Fig.~\ref{band_structure}(a). The wave vectors satisfy $|\bq_{1,2,3}| = 4\pi/\sqrt{3}L$, where $L$ is the periodicity of the potential. We set the Fermi velocity to $v_F=2.55 \eV\rm\AA$, the experimentally measured value in $\text{Bi}_2\text{Te}_3$~\cite{science.1173034,Hsieh2009ATT}. For the moment we neglect higher-order corrections~\cite{LF_2009} to the Dirac cone dispersion~\eqn{model_Hamiltonian}. Then Eq.~\eqn{model_Hamiltonian} depends only on a single dimensionless parameter $w/(v_F/L)$.

The superlattice potential has a profound effect on the Dirac cone. 
The dispersion of the first band above charge neutrality [red line in Fig.~\ref{band_structure}(b)] becomes nearly flat at the $\bK$ and $\bK'$ points of the moir\'e Brillouin zone. Specifically, near $\bK$ and $\bK'$, symmetry constrains the Taylor expansion of the dispersion to the form $\epsilon_\pm(\bk)=\alpha k^2\pm\eta(k^3_x-3k_xk^2_y)$, where $\pm$ indicates valley, expressions for the coefficients $\alpha$ and $\eta$ are given in Appendix~\ref{ap:kp_hot_spots}. This yields three VHSs near $\bK$ and $\bK'$ that map into each other under $C_{3z}$ and provide the large diverging density of states (DOS) shown in Fig.~\ref{band_structure}(c). In addition, there is a local maximum (minimum) for $\alpha<0$ $(>0)$ at $\bk=0$.
At a critical value of $w/(v_F/L)$ where $\alpha=0$, the three nearby VHSs merge to form higher-order VHSs at $\bK$ and $\bK'$ with a power-law diverging DOS~\cite{PhysRevB.101.125120,LFu_HOVHS_2019,PhysRevResearch.2.013355,Daniele_HOVHS_2021}, as found in Ref.~\cite{PhysRevX.11.021024}. Our results do not require fine tuning to the critical value of $w/(v_F/L)$. Instead, we focus on the vicinity of the VHSs that produce a sharp peak in the DOS [highlighted by the horizontal line in Fig.~\ref{band_structure}(b)].

To describe the magnetic instability with momentum $\bK'-\bK$ we consider the electron-electron interaction coupling the hot-spot regions around $\bK$ and $\bK'$. 
Note that screening by low-energy electrons at momentum transfer $\bq=\bK'-\bK$ are weak because electrons at $\bK$ and $\bK'$ occupy orthogonal Bloch states~(see Appendix~\ref{ap:symmetries}).

For simplicity we focus on the Hubbard interaction:
\begin{equation}
\label{local_interaction}
    \hat H_\text{int}=U\int d^2\br\,\hat \Psi^\dagger_\uparrow(\br)\hat \Psi^\dagger_\downarrow(\br)\hat \Psi_\downarrow(\br)\hat \Psi_\uparrow(\br);\quad U>0,
\end{equation}
where $\hat \Psi_{\uparrow,\downarrow}(\br)$ is the electron annihilation operator, $\hat \Psi_\sigma(\br)=\sum_{\bk,\bG} e^{i(\bk-\bG)\br}\,\hat c_{\bk,\bG,\sigma}/\sqrt{A}$ with $A$ area of the sample, $\bk$ and $\bG$ label the Bloch momentum and reciprocal lattice vector of the moir\'e Brillouin zone, respectively, and $\hat c^\dagger_{\bk,\bG,\sigma}$ creates an electron at $\bk-\bG$ with spin $\sigma$. 
Notice that Ref.~\cite{PhysRevX.11.021024} considered an attractive interaction $(U<0)$ and found that the VHS enhances the superconducting critical temperature.

 \subsection{Spin density wave operators}
The interaction term $\hat H_\text{int}$ drives an instability towards density wave ordering, modulated by the nesting vectors, $\bm\delta_{j}$, which connect the $\bK$ and $\bK'$ regions where the DOS diverges at the VHS. The nesting vectors are indicated by green vectors in Fig.~\ref{band_structure}(a),(c) and define a magnetic Brillouin zone [solid cyan line in Fig.~\ref{band_structure}(a)] three times smaller than the original one. 
Spin-momentum locking is expected to drive the formation of an exotic spin-density wave (SDW)~\cite{Chou_2021}. 
To study the SDW, we decompose the order parameters for each modulation $\bm \delta_j$ \cite{LF_2009} into an in-plane direction parallel to the nesting vector, $\bm v_{\parallel j}=\bm e_j\equiv\bm \delta_j/|\delta_j|$; an in-plane direction perpendicular to the nesting vector, $\bm v_{\perp j}=-i(\bz\times \bm e_j)$; and an out-of plane part, $\bm v_{z j}=-i\bz$. 
Correspondingly, for each wave vector $\bm \delta_j$, the SDW operator can be decomposed as
\begin{equation}
\hat S_{\mu j}=\sum_\bk\sum_\bG \hat c^\dagger_{\bk+\bm\delta_j,\bG,\sigma}\,\bm v_{\mu j}\cdot\bm\sigma_{\sigma\sigma'}\,\hat c_{\bk,\bG,\sigma'},
\end{equation}
which represents the three types of magnetic order illustrated in Fig.~\ref{band_structure}(d). We observe that we have $9$ independent spin-density waves corresponding to different combinations of $\bm \delta_j$ and $\bm v_{\mu j}$.
\begin{table}[]
\centering
\begin{tabular}{|c||c|c|c|c|}
\hline
SDW & $C_{3z}$  & $m_{x}$ &  $T$   &  $m_y$  \\
\hline\hline
$\hat S_{\mu 1}$ & $\hat S_{\mu 2}$  & $\hat S^\dagger_{\mu 1}$ &  $-\hat S^\dagger_{\mu 1}$   &  if $z,\parallel:-\hat S_{\mu 1}$;\,else $\hat S_{\perp 1}$     \\
\hline
$\hat S_{\mu 2}$ & $\hat S_{\mu 3}$  & $\hat S^\dagger_{\mu 3}$ &  $-\hat S^\dagger_{\mu 2}$   & if $z,\parallel:-\hat S_{\mu 3}$;\,else $\hat S_{\perp 3}$   \\
\hline
$\hat S_{\mu 3}$& $\hat S_{\mu 1}$  & $\hat S^\dagger_{\mu 2}$ &  $-\hat S^\dagger_{\mu 3}$   &  if $z,\parallel:-\hat S_{\mu 2}$;\,else $\hat S_{\perp 2}$  \\
\hline
\end{tabular}
\caption{Symmetries of the SDW operator $\hat S_{\mu j}$ playing a key role in determining the Ginzburg-Landau free energy. The transformation $m_y$ acts differently on $\mu=\parallel,z$ and $\perp$.}
\label{tab:Symmetries}
\end{table}

\subsection{Symmetries}


The interacting Hamiltonian $\hat H$ composed by $\hat H_0$~\eqn{model_Hamiltonian} and $\hat H_{\rm int}$~\eqn{local_interaction} is invariant under time-reversal $T=\sigma_y K$, where $K$ indicates complex conjugation, as well as a three-fold rotational symmetry, $C_{3z}=e^{i\pi\sigma_z/3}$ and the mirror symmetries, $m_y=i\sigma_y$ and $m_x=i\sigma_x$, which act on both spin and spatial coordinates. Combinations of these imply invariance under two- and six-fold rotation symmetries, $C_{2z}=i\sigma_z$, $C_{6z}=e^{i\pi\sigma_z/6}$, respectively. In order to determine the leading spin-density wave instability it is crucial to notice that the Bloch states $\ket{u_{\bK}}$ and $\ket{u_{\bK'}}$ of the Hamiltonian~\eqn{model_Hamiltonian} at the hot spots are singly degenerate and have opposite mirror $m_y$ eigenvalues. This can be readily understood observing that the time-reversal symmetry $T$ sends $\bK\to\bK'$ and $T m_y T^{-1}=m_y$. Thus, given $\ket{u_{\bK}}$ with $m_y\ket{u_{\bK}}=i\ket{u_{\bK}}$, it follows that $m_y \ket{u_{\bK'}} = m_y T\ket{u_{\bK}}=-iT\ket{u_{\bK}}=-i\ket{u_{\bK'}}$. The potential $w(\br)$ breaks particle-hole symmetry. The action of the symmetries of the model on the spin density wave operators $\hat S_{\mu j}$ is illustrated in Table~\ref{tab:Symmetries}.

\section{Ginzburg-Landau mean field theory}
We study the magnetic order by deriving a Ginzburg-Landau theory. We decouple the local interaction~\eqn{local_interaction} introducing the Hubbard-Stratonovich fields $n(\br,\tau)$ and $\bm m(\br,\tau)$~\cite{Mudry_book,fradkin_2013,altland_simons_2010} for the charge and the magnetization densities to obtain the Lagrangian: 
\begin{equation}
\begin{split}
\label{HS_action}
    &\mathcal L=\int d^2\br \Psi^\dagger\left(\partial_\tau+H_0(\br)-\mu\right)\Psi\\
    &+\frac{U}{2}\int d^2\br\left[  n\,\rho- \bm m\cdot\bm S\right] +\frac{U}{4}\int d^2\br\left[\bm{m}^2-  n^2\right],
\end{split}
\end{equation}
where for the sake of space we omit the dependencies of the fields on the imaginary time $\tau$ and position $\br$, and we have introduced the fermionic operators $\hat\rho(\br)=\sum_\sigma\hat \Psi^\dagger_\sigma(\br)\hat \Psi_\sigma(\br)$ and $\hat S_a(\br)=\bm \hat \Psi^\dagger(\br)\sigma_a\hat \Psi(\br)$. 
Integrating out the electronic degrees of freedom of Eq.~\eqn{HS_action} we obtain the effective action for $n(\tau,\br)$ and $\bm m(\br,\tau)$. 
Then, we take the semiclassical limit of a static order parameter. 
More specifically, we assume a homogenous electron density $n$ and a spatially-modulated magnetization $\bm m(\br)=\sum^3_{j=1}\left(\bm m_{\bm \delta_j}e^{i\bm\delta_j\br}+\bm m_{-\bm\delta_j} e^{-i\bm\delta_j\br}\right)$ where $\bm m_{\bm\delta_j}$ is the Fourier amplitude associated to the modulation $\bm\delta_j$. Since $\bm m(\br)$ is real, $\bm m_{\bm\delta_j}=\bm m^*_{-\bm\delta_j}\equiv\bm m_{j}$. We expand $\bm m_j=\sum_\mu s_{\mu j}\bm v_{\mu j}$ where $\bm v_{\mu j}$ are the normal modes in Fig.~\ref{band_structure}(d) and the order parameter $s_{\mu j}=\langle \hat S_{\mu j}\rangle/A$ is the average value of $\hat S_{\mu j}$ with $A$ area of the sample. 
As a result we find the free energy:
\begin{equation}
    F = \frac{A}{2}\left(\sum_{\mu j}|s_{\mu j}|^2-\frac{n^2}{2}\right)+ \sum_{m=1}^{\infty}\Tr\left[\frac{\left(-G_0\,X\right)^m}{m\,U}\right],\label{free_energy}
\end{equation}
where $G_0=(-\partial_\tau-H_0+\mu)^{-1}$ is the non-interacting single-particle Green's function and $X=U\bm m(\br)\cdot\bm\sigma/2-Un/2$ describes the interaction between the order parameter and the electrons.

\begin{figure}[]
\centering
\includegraphics[width=0.48\textwidth]{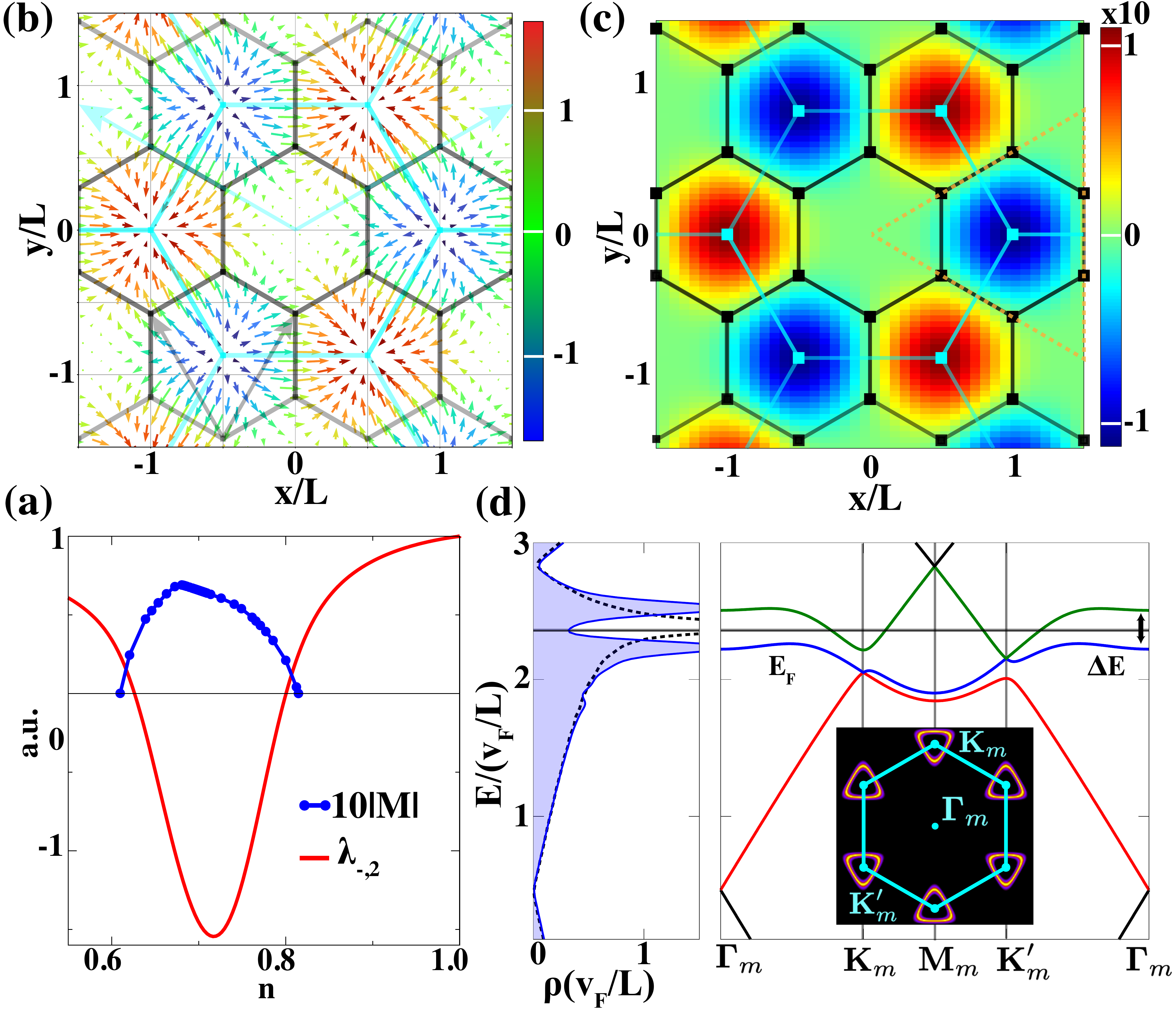}
\vspace{-0.35cm}
\caption{\textbf{Saddle-point solution, real space magnetization and electronic structure in the meron lattice phase.} (a) Blue and red data show the total magnetization $|M|$ and the lowest eigenvalue of $\mathcal M^{\mu\nu}_{jl}$, respectively. (b) Contour plot of the magnetization in real space. Color indicates the out-of-plane component. (c) Winding number $\bm m\cdot(\partial_x\bm m\times \partial_y\bm m)/4\pi$ in real space. 
Black and cyan lines show the superlattice and magnetic unit cells, respectively.
(d) Density of states and electronic dispersion of the meron lattice state. The dashed line shows the density of states of the normal phase. The horizontal line shows the Fermi energy $E_F$; $\Delta E$ denotes the induced gap at $\bm\Gamma_m$. Inset shows the Fermi surface of the magnetic state. The bands are obtained from the Hartree-Fock solution at $w/(v_F/L)\simeq1.54$, $2w=40\meV$, $U=30\,\meV$ and $T=1.5\Kelvin$.}
\label{Sk_el}
\end{figure}

The self-consistency equations are obtained by minimizing the free energy~\eqn{free_energy} with respect to the variational parameters $n$ and $s_{\mu j}$, i.e., $\delta F/\delta n=0$ and $\delta F/\delta s^*_{\mu j}=0$. The magnetic order is determined by finding the roots of these equations numerically, as detailed in Appendix~\ref{ap:saddle_point}.
We find that in the range of filling between $n\sim0.61$ and $n\sim0.8$ the minimum of the free energy~\eqn{free_energy} develops a SDW magnetic ordering. 
The blue data in Fig.~\ref{Sk_el}(a) shows the magnitude of the total magnetization density, $|M|=\sqrt{\sum_{\mu j }|s_{\mu j}|^2}$, as a function of the filling per unit cell of the superlattice.

In real space, the magnetic order $\bm m(\br)$ obtained from the saddle-point solution forms a N\'eel-type meron lattice~\cite{Nagaosa_2013} with two merons in each unit cell, as shown in Fig.~\ref{Sk_el}(b). The merons are well-defined because the magnetic moment is forced to be in-plane along mirror-invariant lines. Each meron can be characterized by the Pontryagin density $\Phi(\br)=\,\hat{\bm m}(\br)\cdot\left[\partial_x\hat{\bm m}(\br)\times\partial_y\hat{\bm m}(\br)\right]/4\pi$ shown in Fig.~\ref{Sk_el}(c) where $\hat{\bm m}$ is a unit magnetization vector. 
We find $\int_S d^2\br \Phi(\br) = \pm 1/2$, where $S$ is the triangular domain illustrated by the orange dashed line in Fig.~\ref{Sk_el}(c).  
In the center of each unit cell is an intermediate region of destructive interference with vanishing magnetization. The magnetic order spontaneously breaks the translation symmetry of the potential, giving rise to a magnetic unit cell [cyan line in Fig.~\ref{Sk_el}(b)] three times larger ($\sqrt{3}\times\sqrt{3}$) than the original one [black solid line in Fig.~\ref{Sk_el}(b)]. 

\subsection{Hartree-Fock Hamiltonian}

In this section we discuss the properties of the electronic band structure in the magnetic phase. 
We treat the effect of the magnetization on the electronic spectrum at the Hartree-Fock level by replacing $H_0(\br)$ in Eq.~\eqref{model_Hamiltonian} with: 
\begin{equation}
\label{Hartree_Fock}
    H_{\text{HF}}(\br)=H_0(\br)-\frac{U}{2}\bm m(\br)\cdot\bm\sigma.
\end{equation}
Despite breaking time-reversal symmetry, the resulting electronic spectrum remains gapless because the magnetic order preserves $C_{2z}T$ symmetry. The dispersion is plotted in the right panel of Fig.~\ref{Sk_el}(d) in the magnetic Brillouin zone: the original red band in Fig.~\ref{band_structure}(b) is decomposed into three different bands given by the red, blue and green lines in Fig.~\ref{Sk_el}(d). The Dirac cones at $\bK_m$ between the red and blue bands and at $\bK'_m$ between the blue and the green ones are protected by $C_{2z}T$. 
Thus, they give rise to a gapless electronic spectrum whose Fermi surface is shown in the inset of Fig.~\ref{Sk_el}(d).

Although the order parameter $\bm m(\br)$ does not open a full gap, it significantly reduces the DOS at the Fermi level (horizontal black dashed line in Fig.~\ref{Sk_el}(d)) by opening a gap between the blue and green bands at $\Gamma_m$ of order $\Delta E\propto U|M|$.
The gap splits the peak in the DOS resulting from the VHS into two peaks above and below, as shown in the left panel of Fig.~\ref{Sk_el}(d). The significant decrease in the kinetic energy from splitting the large peak in the DOS makes the magnetic state energetically favorable with respect to the normal one.

In the following we discuss the symmetries of the magnetic phase and we determine the region of stability of the spin density wave order. 

\section{Phase diagram}

The instability of the Dirac cone surface state is determined by expanding the free energy for small values of the order parameter $s_{\mu j}$. As time-reversal symmetry acts on the order parameter by $T:s_{\mu j}\to-s^*_{\mu j}$ [see Table~\ref{tab:Symmetries}], only even powers of $s_{\mu j}$ are allowed in the free energy~\eqn{free_energy}, so that to second order in $s_{\mu j}$ at fixed density $n$,  $F_{(2)}=\sum_{\mu\nu}\sum^3_{jl=1}s^*_{\mu j}\mathcal{M}^{\mu\nu}_{jl}s_{\nu l}/2$, 
where 
\begin{equation}
\label{effective_mass}
    \mathcal{M}^{\mu\nu}_{jl} = \delta_{\mu\nu}\delta_{jl}-\frac{U}{2}\chi^{\mu\nu}_{jl},
\end{equation}
and $\chi^{\mu\nu}_{jl}\equiv\int^\beta_0d\tau\langle \hat S^\dagger_{\mu j}(\tau) \hat S_{\nu l}(0)\rangle/A$ is the thermodynamic spin susceptibility with $\tau$ imaginary time and $\beta$ inverse temperature. 
The matrix $\mathcal{M}^{\mu\nu}_{jl}$ contains both diagonal terms ($j=l$), which come from momentum-conserving scattering processes, and off-diagonal terms ($j\neq l$) from Umklapp processes with momentum $\bm\delta_j-\bm\delta_l\in\bG$ with $\bG$ a reciprocal lattice vector. The matrix $\mathcal M^{\mu\nu}_{jl}$ is constrained by the symmetries of the Hamiltonian as discussed in Appendix~\ref{ap:symmetries}.

\begin{figure}[]
\centering
\includegraphics[width=0.47\textwidth]{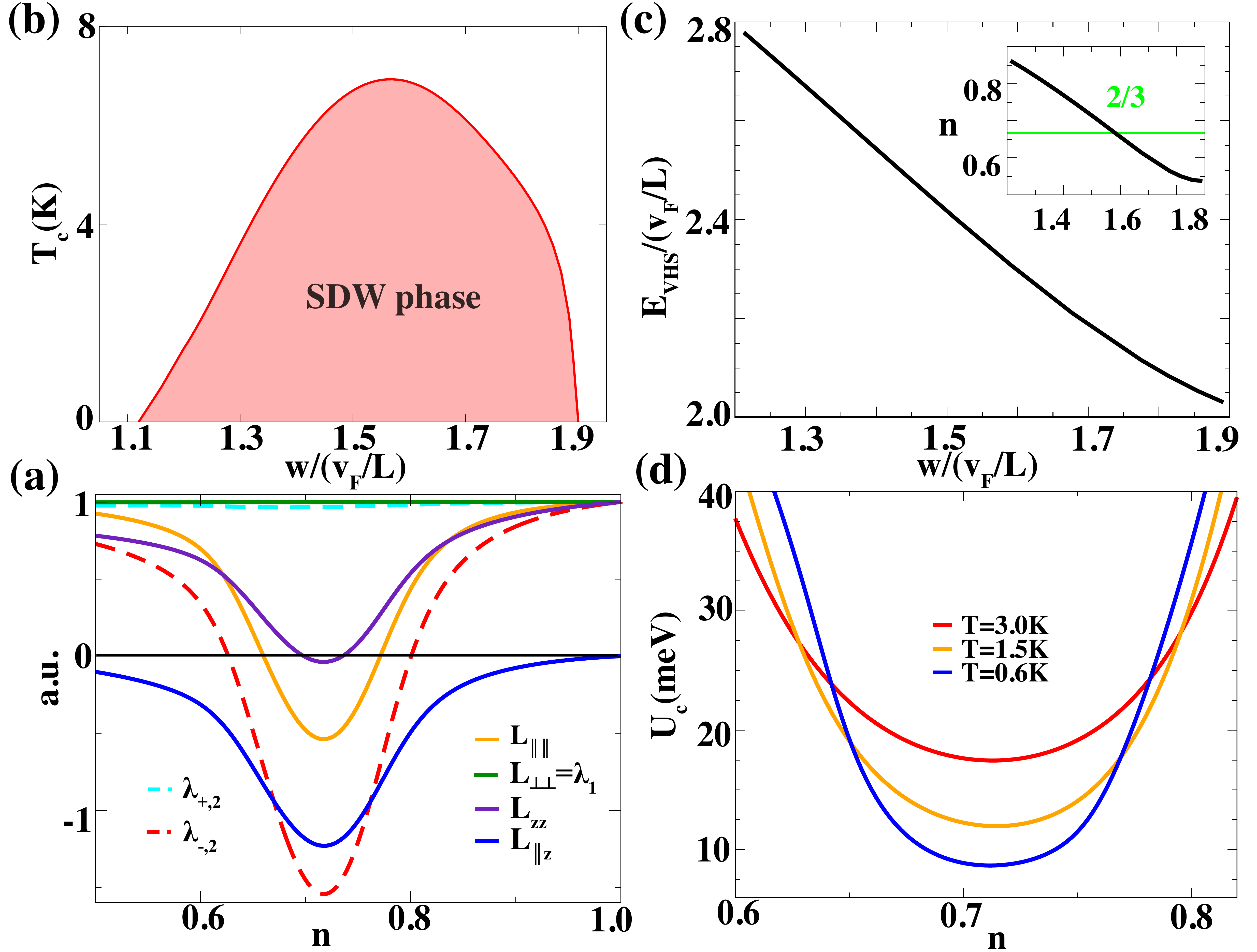}
\vspace{-0.35cm}
\caption{\textbf{Magnetic instability, critical temperature and interaction}. (a) Solid lines show the components of the second order free energy $L_{\mu\nu}$. Red and cyan dashed lines illustrate the eigenvalues $\lambda_{-,2}$ and $\lambda_{+,2}$, respectively, while $\lambda_1=L_{\perp\perp}$ is the solid green line. The calculations were performed at $T=1.5\Kelvin$, $U=30\meV$ and $w/(v_F/L)\simeq1.54$. (b) Critical temperature for the magnetic transition as a function of $w/(v_F/L)$ for $U=30\meV$. For a given $w/(v_F/L)$, the filling $n$ is set so that the Fermi level is at the energy of the VHS. (c) Evolution of the energy of the VHS singularity as a function of $w/(v_F/L)$. Inset shows the electron filling per superlattice unit cell at the VHS versus $w/(v_F/L)$. (d) Critical interaction $U_c$ as a function of the filling $n$ at several temperatures, fixing $2w=40\meV$ and $w/(v_F/L)\simeq1.54$. 
}
\label{LG_HF_results}
\end{figure}

An instability exists when one of the eigenvalues of the matrix $\mathcal{M}^{\mu\nu}_{jl}$ becomes negative, which comprises a generalization of the Stoner criterion~\cite{Stoner_1938}.
The corresponding eigenvector indicates the magnetic configuration of the instability and is classified by how it transforms under symmetry. 
Since the saddle-point solution is $C_{3z}$-symmetric, we consider a $C_{3z}$-invariant eigenvector of $\mathcal{M}^{\mu\nu}_{jl}$, which implies the magnetization takes the form $s_{\mu j} = M U_\mu /\sqrt{3}$. 
It follows that
\begin{equation}
    F_{(2)}=\frac{|M|^2}{2}\left(\begin{matrix}U_\parallel\\U_\perp\\U_z\end{matrix}\right)^\dagger\left(\begin{matrix} L_{\parallel\parallel} & 0 & L_{\parallel z} \\0 & L_{\perp\perp}  & 0 \\ L_{\parallel z} & 0 & L_{z z} \end{matrix}\right)\left(\begin{matrix}U_\parallel\\U_\perp\\U_z\end{matrix}\right),
    \label{eq:F2}
\end{equation}
where the elements $L_{\mu\nu} = \sum_{jl} \mathcal{M}^{\mu\nu}_{jl}/3$ are shown in Fig.~\ref{LG_HF_results}(a) as a function of the density at $T=1.5$K. The perpendicular ($U_\perp$) component of the magnetic order is decoupled from the other components ($U_\parallel,U_z$) because it is even under the mirror $m_y$, while the parallel and $z$-components are odd. 
This property can be readily understood by looking at the last column of Table~\ref{tab:Symmetries}, where we show the action of the symmetry $m_y$ on the different components of the spin-density wave order.
From Eq.~(\ref{eq:F2}), the surface Dirac cone is unstable when either $\lambda_1=L_{\perp\perp}<0$, which corresponds to in-plane order purely in the ``spiral-xy'' channel and even under $m_y$, or when one of the eigenvalues $\lambda_{\pm,2}=(L_{\parallel\parallel} + L_{zz})/2 \pm \sqrt{(L_{\parallel\parallel} - L_{zz} )^2/4 + L_{\parallel z}^2} < 0$, which corresponds to a generically non-coplanar SDW with ``120-xy'' and $z$ components, and breaking $m_y$ [see Fig.~\ref{band_structure}(d)]. 
In order to characterize the magnetic instability we compute the spin-susceptibility introduced in Eq.~\eqn{effective_mass}:
\begin{equation}
\begin{split}
    \chi^{\mu\nu}_{jl}=&-\frac{T}{A}\sum_{\bk ,\epsilon_n}\Tr\Big[G_0 (\bk,i\epsilon_n)\,O^\dagger_{\mu j}\,V^{\bm\delta_j-\bm\delta_l}\\
    & G_0 (\bk+\bm\delta_l,i\epsilon_n)\,O_{\nu l}\Big],
\end{split}
\end{equation}
where $G_0(\bk,i\epsilon_n)$ is the single particle Green's function
\begin{equation}
    G_0(\bk,i\epsilon)=\sum_n\frac{\ket{u_{n\bk}}\bra{u_{n\bk}}}{i\epsilon-\xi_{n\bk}},
\end{equation}
where $\ket{u_{n\bk}}$ are the Bloch states of $H_0$, for simplicity we have introduced the operator $O_{\mu j}=\bm v_{\mu j}\cdot\bm \sigma$ and $V^{\bG}$ is the sewing matrix $V^{\bG}_{\bQ,\bQ^\prime}=\delta_{\bQ,\bQ^\prime+\bG}$.
We find that for a range of filling near $2/3$ electron per unit cell, the dominant instability occurs in the channel corresponding to $\lambda_{-,2}$, as shown by the region where the red solid line in Fig.~\ref{Sk_el}(a) goes negative, and in agreement with our saddle-point solution. This solution, which breaks $m_y$, is energetically favored over the in-plane spin solution that is even in $m_y$ because it allows a gap to open at $\bm \Gamma_m$ in the magnetic BZ, indicated by $\Delta E$ in Fig.~\ref{Sk_el}(d).
Indeed, the latter solution being odd under $m_y$ allows mixing between the Bloch states $\ket{u_{\bK}}$ and $\ket{u_{\bK'}}$ at the hot spot regions which are characterized by opposite mirror $m_y$ eigenvalues as detailed in Appendix~\ref{ap:symmetries}.
The resulting magnetic state is a pattern of half-integer topological vortices, with the same winding and opposite polarities corresponding to a meron lattice.
The other magnetic state, with eigenvalue $\lambda_{+,2}$, also breaks $m_y$, but is energetically unfavourable because it exhibits a clockwise magnetization winding inconsistent with the intrinsic SOC of the surface state in Eq.~\eqn{model_Hamiltonian}. We have also studied the $C_{3z}$-breaking magnetic orders. Since these orders do not hybridize the states $\ket{u_{\bK}}$ and $\ket{u_{\bK'}}$, they do not lead to a magnetic instability. 
The symmetry of the magnetic orders are further discussed in Appendix~\ref{ap:symmetries}.



The critical temperature and interaction strength occur at precisely the point where one of the eigenvalues of $\mathcal M^{\mu\nu}_{jl}$ changes sign. Fig.~\ref{LG_HF_results}(b) shows the critical temperature for a range of fillings around the VHS at $U=30$meV. The magnetic dome peaks at $T_c\sim 7$K near an optimal value of $w/(v_F/L)\simeq1.54$: surprisingly, the peak does not coincide with the higher-order VHS \cite{PhysRevX.11.021024} at $w/(v_F/L) \simeq 1.36$. Fig.~\ref{LG_HF_results}(c) shows the evolution of the energy and the number of electrons per superlattice unit cell where the VHS occurs as a function of $w/(v_F/L)$. The maximum critical temperature for the magnetic transition takes place around $n\sim2/3$ filling, denoted as the horizontal green line in the inset of Fig.~\ref{LG_HF_results}(c).

Fig.~\ref{LG_HF_results}(d) shows the critical interaction $U_c$ necessary to induce the magnetic instability as a function of the filling $n$ for different temperatures $T$. Lowering $T$ reduces the magnetic transition to arbitrarily weak repulsive interactions. 
In contrast, in the absence of the superlattice potential, spontaneous magnetization of the surface state~\cite{LF_2009,2012_1_AStern,2012_2_AStern,Wu_Si_2011,PhysRevB.86.155146,PhysRevB.86.161110,PhysRevB.88.205107,PhysRevB.91.155405} in the  $\rm Bi_{2}\rm Se_{3}$ family requires a critical value of the Coulomb interaction, on the order of $2$eV~\cite{2012_1_AStern,2012_2_AStern,PhysRevB.91.155405}. 
We expect our results are robust to weak disorder, which will reduce the transition temperature~\cite{Litak_1998,Wilson_2020,PhysRevB.105.075144} without eliminating the magnetic state.

\subsection{Hexagonal warping}

Beyond the linear momentum dependence of the Dirac cone~\eqref{model_Hamiltonian} the dispersion develops a hexagonal warping term $H_w(\bk)=\lambda(k^3_x-3k_xk^2_y)\sigma_z$~\cite{LF_2009}. Despite being small in $\lambda/v_FL^2$ this correction gives rise to a series of interesting effects that are experimentally relevant. 
The hexagonal warping breaks $m_y$ and $C_{2z}T$ symmetries which gives a finite $U_\perp$ component in the magnetization pattern. 
While the average magnetic moment in the unit cell continues to vanish, $\int d^2\br ~m_i=0$, the explicitly broken $m_y$ symmetry gives rise to a finite out-of-plane toroidal moment $\mathcal T_z= \int d^2\br\,(x m_y-ym_x)/2$~\cite{PhysRevLett.95.237402,PhysRevB.76.214404,bhowal2022magnetoelectric} which typically manifests in the magnetoelectric susceptibility $\alpha_{xy}=-\alpha_{yx}$~\cite{Shudan_2016,PhysRevB.92.235205}.


\section{Conclusions} 

We have shown that a superlattice potential on the surface of a TI provides a new route to spontaneously breaking time reversal symmetry on the TI surface. 
The magnetic order realizes a meron lattice exhibiting pairs of merons with opposite topological charge in the unit cell. 
The periodicity of the meron lattice is determined by the period of the potential; consequently, the meron lattice periodicity is completely tunable. Although the magnetic order breaks time-reversal, it preserves $C_{2z}T$ and thus does not open a gap on the TI surface. 

The meron lattice phase can be measured by imaging the magnetization in real space through Lorentz transmission electron microscopy~\cite{Yu_2010,Heinze_2011} and nitrogen vacancy magnetometry~\cite{Dovzhenko2018} or in reciprocal space via X-ray diffraction~\cite{Rosch2009,PhysRevB.76.224424,PhysRevB.85.220406,PhysRevLett.102.037204,PhysRevLett.107.127203,PhysRevLett.102.037204,Hirschberger2019SkyrmionPA,Brearton_2021}. The magnetization can also be measured through the magneto-optical Kerr effect~\cite{MOKE_2017,Liu_2020} and reflective magnetic circular dichroism~\cite{RMCD_2018}.
In addition, the magnetoelectric susceptibility $\alpha_{xy}$ is observed either by measuring an in-plane magnetization in response to an electric field, or a current resulting from an in-plane magnetic field.
The reduced density of states resulting from the magnetic order gives a drastic variation of the electronic compressibility across the transition. Finally, we also expect that the response to a magnetic field~\cite{PhysRevLett.83.3737,Taguchi_2001,PhysRevLett.93.096806,Nagaosa_2004,Binz_2008,Nagaosa_2012,Rosch_2012} gives rise to distinctive signatures of the meron state.  

This finding has far-reaching consequences, as this unconventional magnetic state will have implications both on the experimental and on the theoretical level. First, the topological spin texture realizes an electromagnetic field on the scale of the superlattice that can be employed by proximity effects to modify band structure and topological properties of electronic systems~\cite{Hurst2015,Shimizu_2021,guan2021unconventional,Paul_2021,divic2021magnetic,MillisStaggeredField21,MillisHartreeFockTMD21}.
Second, the broken mirror symmetry that results from hexagonal warping of the Fermi surface implies that the meron lattice might host a magnetoelectric response with potential application to spintronics~\cite{AFert_2013,AFert_2017,Zhang_2015,Zhang_2020,MillisStaggeredField21,MillisHartreeFockTMD21}. This effect can be enhanced by strain~\cite{Bi_2019,Balents_2019,He_2020} and lattice relaxation~\cite{Koshino_2017}. Finally, the interplay between the spin density wave ordering and possible superconducting instabilities~\cite{PhysRevX.11.021024} is an open problem which is left to future studies. 


\begin{acknowledgments}
\emph{Acknowledgements.---} We have benefited from discussions with Yang-Zhi Chou, M. Michael Denner, Shiang Fang, Jiawei Zang, Andrew J. Millis, Zhentao Wang, Tiancheng Song, and Justin Wilson. We are grateful to Lucy Reading-Ikkanda for creating the figure of the experimental setup and the sketch of the magnetic state. We also acknowledge the support of the Flatiron Institute, a division of the Simons Foundation. This work was partially supported by the Air Force Office of Scientific Research under Grant No. FA9550-20-1-0260 (J.C.) and Grant No.~FA9550-20-1-0136 (J.H.P.) and the Alfred P. Sloan Foundation through a Sloan Research Fellowship (J.H.P.). J.H.P. and J.C. acknowledge hospitality of the Aspen Center for Physics, where some of this work was developed and which is supported by National Science Foundation grant PHY1607611.
\end{acknowledgments}

\appendix

\section{The $\bk\cdot\bp$ expansion around $\bK$ and $\bK'$}
\label{ap:kp_hot_spots}

 The origin of the high order van Hove singularities (HOVHS)~\cite{PhysRevB.101.125120,LFu_HOVHS_2019,PhysRevResearch.2.013355,Daniele_HOVHS_2021} can be derived from the lattice symmetries of the Hamiltonian $H_0(\br)$. We define the small deviation from the $\bK$ point, $\bq=\bk-\bK$, and the linear combinations $q_{\pm}=q_x\pm iq_y$, which transforms under $C_{3z}$ as $q_{\pm}\to e^{\pm i2\pi/3}q_{\pm}$. At second order in $\bq$ the only allowed term in the energy dispersion is $q_+q_-=q^2$. The next contributions invariant under $C_{3z}$ are $q^3_+$ and $q^3_-$. As a consequence of the mirror $m_y$ symmetry the cubic term is $(q^3_++q^3_-)/2=q^3_x-3q^2_yq_x$. 
\begin{figure}
    \centering
    \includegraphics[width=0.28\textwidth]{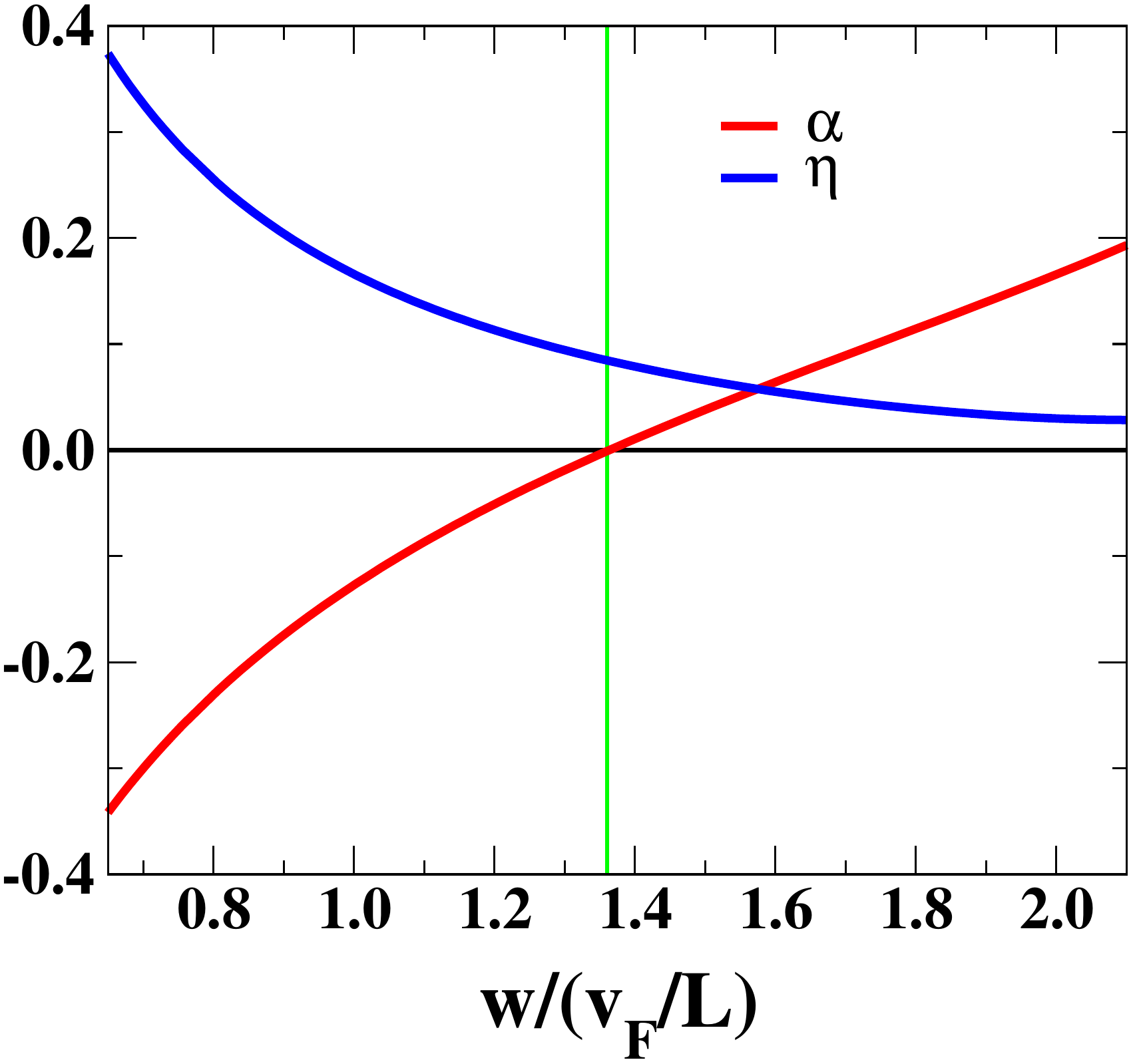}
    \vspace{-0.3cm}
    \caption{Parameters $\alpha$ and $\eta$ of the $\bk\cdot\bp$ expansion of the dispersion relation around $\bK$. The vanishing of $\alpha$ at $w/(v_F/L)=1.36$ highlighted by the green line implies a HOVHS.}
    \label{kp_HOVHS}
\end{figure}
Thus, to cubic order in $\bq$ the dispersion around $\bK$ reads~\cite{PhysRevX.11.021024} $\epsilon_{\bK}(\bq)=\epsilon_0+\alpha q^2+\eta(q^3_x-3q_xq^2_y)+\cdots$. 
Applying time-reversal symmetry we find that the expansion at $\bK'$ reads $\epsilon_{\bK'}(\bq)\simeq\epsilon_0+\alpha q^2-\eta(q^3_x-3q_xq^2_y)$.
The values of the coefficients $\alpha$ and $\eta$ are obtained from the $\bk\cdot\bp$ perturbation theory at $\bK$. 
Given $H_0(\bK+\bq)-H_0(\bK)$ expanding to the third order in the small deviation $\bq=\bk-\bK$ yields:
\begin{equation}
\label{kp_alpha}
    \alpha=2\sum_{m\neq n}\frac{\Re\left[\mel{u_{n\bK}}{\sigma_+}{u_{m\bm K}}\mel{u_{m\bK}}{\sigma_-}{u_{n\bK}}\right]}{\epsilon_{n\bK}-\epsilon_{m\bK}},
\end{equation}
and
\begin{equation}
\begin{split}
\label{kp_beta}
    \eta=&\sum_{m,j\neq n}2\Im\Bigg[\frac{\mel{u_{n\bK}}{\sigma_-}{u_{m\bK}}\mel{u_{m\bK}}{\sigma_-}{u_{j\bK}}}{(\epsilon_{n\bK}-\epsilon_{m\bK})(\epsilon_{n\bK}-\epsilon_{j\bK})}\\
    &\mel{u_{j\bK}}{\sigma_-}{u_{n\bK}}\Bigg],
\end{split}
\end{equation}
where $\ket{u_{n\bK}}$ and $\epsilon_{n\bK}$ are the Bloch state and the eigenvalue, respectively, of $H_0(\br)$ at $\bK$, $n$ is the first positive energy band and $\sigma_{\pm}=\sigma_x\pm i\sigma_y$. The evolution of the couplings $\alpha$ and $\eta$ is shown in Fig.~\ref{kp_HOVHS}.
The higher-order van Hove singularity occurs when the quadratic term $\alpha$ in Eq.~\eqn{kp_alpha} vanishes at $w/(v_F/L)\simeq1.36$, highlighted by the vertical green line in Fig.~\ref{kp_HOVHS}. The result is a power-law divergence in the density of states $\rho(\epsilon)\sim|\epsilon|^{-1/3}$. 
By looking at the solutions of $\nabla_\bq \epsilon_\pm(\bq)=0$ ($\pm$ for $\bK$ and $\bK'$ respectively), we find that $\bq=0$ is a maximum for $\alpha<0$, a minimum for $\alpha>0$ and, finally, a higher-order critical point for $\alpha=0$. Away from the origin there are three further solutions at $\bm \kappa_{\pm,j}=\mp2\alpha C^{j-1}_{3z}(1,0)/3\eta$, which are saddle points. Approaching the higher-order VHS at $\alpha\to0$, these three saddle points merge at $\bq=0$, {\it i.e.}, at $\bK$ or $\bK'$.

\section{Solution of the saddle-point equations} 
\label{ap:saddle_point}

In this section we detail the mean-field solution of the interacting electrons on the surface of the TI subject to the superlattice potential $w(\br)$.
Minimizing $F$~\eqn{free_energy} with respect to the filling $n$ and the magnetic configurations $\bm m(\br)$ we find the saddle-point equations: 
\begin{equation}
\label{filling}
    \frac{\delta F}{\delta n}=0\Rightarrow n=\frac{1}{A}\Tr\left[\left(G_0^{-1}+X\right)^{-1}\right],
\end{equation}
and
\begin{equation}
\label{spin_ordering}
    \frac{\delta F}{\delta s^*_{\mu j}}=0\Rightarrow s_{\mu j}=\frac{2}{UA}\Tr\left[\frac{\delta X}{\delta s^*_{\mu j}}\left(G^{-1}_0+X\right)^{-1}\right].
\end{equation}
In the main text we show the result of the numerical solution of Eqs.~\eqn{filling} and~\eqn{spin_ordering}, which was obtained as follows. Eqs.~\eqn{filling} and \eqn{spin_ordering} take a simple form in the basis of the eigenstates $\ket{\phi_{n\bk}}$ of the Hartree-Fock Hamiltonian in Eq.~\eqn{Hartree_Fock} with eigenvalues $\bar\epsilon_{n\bk}$:
\begin{equation}
\label{filling_1}
     n=\frac{1}{3N}\sum_{n\bk} f(\bar\xi_{n\bk})-n_0,
\end{equation}
where $\bar\xi_{n\bk}=\bar\epsilon_{n\bk}-\mu^*$ and $\mu^*=\mu-Un/2$;
and
\begin{equation}
\label{spin_ordering_1}
   s_{\mu j}=\frac{1}{3N}\sum_{n\bk}f(\bar\xi_{n\bk})\mel{\phi_{n\bk}}{O^\dagger_{\mu j}V^{\bm\delta_j}}{\phi_{n\bk}},
\end{equation}
where the filling $n$ is measured with respect to the charge neutrality point $n_0$, $N$ is the number of points sampling the Brillouin zone, the factor of three takes into account the size difference between the original and the magnetic Brillouin zone. Moreover, the function $f(\epsilon)=1/(e^{\beta\epsilon}+1)$ is the Fermi-Dirac distribution function, and $V^{\bG}$ is the sewing matrix satisfying $V^{\bG}_{\bQ,\bQ^\prime}=\delta_{\bQ,\bQ^\prime+\bG}$, $V^{\bG}\ket{\phi_{n\bk}}=\ket{\phi_{n\bk+\bG}}$. The self-consistent equations is performed by a find-root algorithm.

\section{Symmetries of the free energy and classification of magnetic orderings} 
\label{ap:symmetries}

In this Appendix we discuss the symmetries of the second order matrix $\mathcal M^{\mu\nu}_{jl}$ and classify the different magnetic orderings according to their symmetry properties. 
The second order tensor is constrained by the symmetries in Table~\ref{tab:Symmetries}.
The three-fold rotation about the $z$-axis, $C_{3z}:s_{\mu j}\to s_{\mu j+1}$, gives $\mathcal{M}^{\mu\nu}_{jl}=\mathcal{M}^{\mu\nu}_{j+1l+1}$. As a result  $\mathcal{M}^{\mu\nu}_{jl}$ is expressed in terms of two three-dimensional matrices: $\Lambda_{\mu\nu}=\mathcal{M}^{\mu\nu}_{j,j}$, $\Omega_{\mu\nu}=\mathcal{M}^{\mu\nu}_{j,j+1}$ while $\mathcal{M}^{\mu\nu}_{j+1,j}=\Omega^*_{\nu\mu}$ since the free energy $F_{(2)}$ is real. In addition, the mirror symmetry $m_x$~[given in Table \eqn{tab:Symmetries}] implies that $\Lambda_{\mu\nu}$ is real and symmetric, $\Omega_{\mu\nu}=\Omega_{\nu\mu}$ but with complex elements $\Omega^*_{\mu\nu}\neq\Omega_{\nu\mu}$. Finally, the remaining $m_y$ symmetry imposes that the matrix elements $\Lambda_{\parallel\perp}=0$ and $\Lambda_{\perp z}=0$, $\Omega_{\parallel \perp}$ and $\Omega_{\perp z}$ are purely imaginary, while the other components of $\Omega_{\mu\nu}$ are reals. The latter constraint implies that the $C_{3z}$ symmetric spiral-xy order decouples from the 120-xy and out-of-plane orderings.
As a result by projecting the second order free energy $F_{(2)}$ on the $C_{3z}$ symmetric configuration $s_{\mu j}=\xi\,U_\mu/\sqrt{3}$ we find the expression in the main text Eq.~\eqn{eq:F2} where $L_{\mu\nu}=\Lambda_{\mu\nu}+2\Re\Omega_{\mu\nu}$.

\begin{table}
\begin{center}
\begin{tabular}{ |c|c|c|c| } 
 \hline
  & $E$ \quad&\quad 2$C_3$ \quad&\quad 3$m$ \quad\\ 
 \hline\hline
 $\quad \Gamma_1\quad$  &\quad  1 \quad &\quad  1\quad  &\quad  1\quad  \\ 
 \hline
 $\quad \Gamma_2\quad $ & \quad 1 \quad &\quad  1\quad  &\quad  -1\quad  \\
 \hline
 $\quad \Gamma_3\quad $ & \quad 2\quad  & \quad -1\quad  & \quad 0\quad  \\
 \hline
\end{tabular}
\caption {Character table of $C_{3v}$. $E$, $C_3$ and $m$ represent the conjugation classes of the identity, $C_{3z}$ and $m_y$, respectively.} \label{character_table} 
\end{center}
\end{table}

We now characterize the spectrum of $\mathcal M^{\mu\nu}_{jl}$ and we classify the magnetic states with respect to their symmetry behaviour under the point-group symmetries of the model. 
The magnetic ordering is described by the real order parameter $\bm m(\br)=\sum_{j}\bm m_{j}\,e^{i\bm\delta_j\cdot\br}+c.c.$ which is completely determined by the Fourier amplitudes $\bm m_j=\sum_\mu s_{\mu j}\bm v_{\mu j}$.
The nine magnetic configurations originating from $s_{\mu j}$ are decomposed in one one-dimensional representation $\Gamma_1$, two one-dimensional representations $\Gamma_2$ and three two-dimensional representations $\Gamma_3$ given in Table~\ref{character_table}. 
The one-dimensional representations $\Gamma_1$ and $\Gamma_2$ are symmetric under $C_{3z}$ and corresponds to equal superpositions of the modulations $\bm \delta_{1,2,3}$. The parity of $\Gamma_1$ under $m_y$ is even consistently with the spiral-xy order. On the other hand, $\Gamma_2$ is odd under $m_y$ like the 120-xy and out-of-plane orderings. Due to the intrinsic spin-orbit coupling in-plane 120-xy and out-of-plane orderings mix to form the meron lattice state. Finally, we have the two-dimensional representation $\Gamma_3$. Here, we find two eigenstates of $C_{3z}$ characterized by eigenvalues $\omega=e^{2\pi i/3}$ and $\omega^*$ and $m_y$ is off-diagonal in this basis. Two dimensional degeneracy implies from this symmetry.

In the normal state the two valleys $\bK$ and $\bK'$ are equally occupied giving rise to a time-reversal symmetric state with vanishing magnetization. 
It is important to notice that the two valleys are characterized by non-degenerate Bloch states, $\ket{u_{\bK}}$ and $\ket{u_{\bK'}}$, [first band above charge neutrality] with opposite mirror eigenvalues. 
The meron lattice state belonging to $\Gamma_2$  breaks the mirror symmetry $m_y$ of $\bK$ and $\bK'$ and couples $\ket{u_{\bK}}$ and $\ket{u_{\bK'}}$. 
The resulting Hartree-Fock orbitals at $\bm \Gamma_m$ of the magnetic \mr Brillouin zone are eigenstates of the mirror symmetry $m_x$ and correspond to a coherent superposition of the two valleys $\bK$ and $\bK'$.

\bibliography{biblio.bib}

\end{document}